# Realizing the Willis equations with pre-stresses

Z.H. Xiang[*], R.W. Yao

*AML, Department of Engineering Mechanics, Tsinghua University, Beijing 100084, China*

**ABSTRACT**

This paper proves that the linear elastic behavior of the material with inhomogeneous pre-stresses can be described by the Willis equations. In this case, the additional terms in the Willis equations, compared with the classical linear elastic equations for homogeneous media, are related to the gradient of pre-stresses. In this way, the material length scale is naturally incorporated in the framework of continuum mechanics. All these findings also coincide with the results of transformation elastodynamics, so that they can meet the requirement of the principle of material objectivity and the principle of general invariance.

## 1. Introduction

The classical linear elastodynamic theory was established for homogeneous media. In this theory, the elastic wave propagation is described by a constitutive equation

$$\boldsymbol{\sigma} = \boldsymbol{C} : \nabla \boldsymbol{u}, \tag{1a}$$

and an equation of motion

$$\nabla \cdot \boldsymbol{\sigma} + \boldsymbol{f} = \rho \ddot{\boldsymbol{u}}, \tag{1b}$$

where $\boldsymbol{\sigma}$ is the stress tensor; $\boldsymbol{C}$ is the elasticity tensor; $\boldsymbol{f}$ is the body force vector; $\rho$ is the mass density; $\boldsymbol{u}$ is the displacement vector; and the superposed dot denotes the differentiation with respect to time $t$.

---

[*] Corresponding author. Tel: +86-10-62796873.
*Email addresses:* xiangzhihai@tsinghua.edu.cn (Z.H. Xiang), yrw11@mails.tsinghua.edu.cn (R.W. Yao).

The elastodynamic equations for the linear elastic wave propagation in inhomogeneous media were established by Professor Willis over 30 years ago, based on the variational principle and the perturbation theory (Willis, 1981, 1997). The Willis equations are

$$\langle \boldsymbol{\sigma} \rangle = \boldsymbol{C}_{eff} * \langle \boldsymbol{e} \rangle + \boldsymbol{S}_{eff} * \langle \dot{\boldsymbol{u}} \rangle, \tag{2a}$$

$$\nabla \cdot \langle \boldsymbol{\sigma} \rangle + \boldsymbol{f} = \langle \dot{\boldsymbol{p}} \rangle, \tag{2b}$$

$$\langle \boldsymbol{p} \rangle = \boldsymbol{S}_{eff}^{\dagger} * \langle \boldsymbol{e} \rangle + \boldsymbol{\rho}_{eff} * \langle \dot{\boldsymbol{u}} \rangle, \tag{2c}$$

where $\langle \ \rangle$ denotes the ensemble average; $\boldsymbol{e}$ is the strain tensor, $\langle \boldsymbol{e} \rangle = (\nabla \langle \boldsymbol{u} \rangle + \langle \boldsymbol{u} \rangle \nabla)/2$; $\boldsymbol{p}$ is the momentum density; $\boldsymbol{C}_{eff}$, $\boldsymbol{S}_{eff}$ and $\boldsymbol{\rho}_{eff}$ are non-local operators depending on the angular frequency of oscillation; $\boldsymbol{S}_{eff}^{\dagger}$ is the adjoint of $\boldsymbol{S}_{eff}$; and $*$ denotes the time convolution. In contrast to classical elastodynamic equations, the Willis equations can give more accurate predictions of wave behaviors in periodically inhomogeneous media (Willis, 2011; Norris et al., 2012; Srivastava and Nemat-Nasser, 2012). However, they have not been widely implemented for more general cases (Nassar et al., 2015), probably due to the unclear physical meaning and the abstract formulation of $\boldsymbol{S}_{eff}$.

In recent years, transformation optics (Leonhardt, 2006; Pendry et al., 2006) and transformation acoustics (Chen and Chan, 2010) have been widely used to obtain the inhomogeneously distributed effective material parameters of metamaterials (Shamonina and Solymar, 2007), which can steer the wave propagation along a desired trajectory. The basis of these transformation methods is the form-invariance of wave equations under general coordinate transformations. Unfortunately, Milton et al. (2006) pointed out that the classical elastodynamic equations are generally not form-invariant. And amazingly, if taking the deformation gradient as the gauge between the displacements before and after the transformation, one can obtain the Willis equations in frequency domain:

$$\boldsymbol{\Sigma} = \boldsymbol{C} : \nabla \boldsymbol{U} + \boldsymbol{S} \cdot \boldsymbol{U}, \tag{3a}$$

$$\nabla \cdot \boldsymbol{\Sigma} = \boldsymbol{S}^{\mathrm{T}} : \nabla \boldsymbol{U} - \omega^2 \boldsymbol{\rho}_{\mathit{eff}} \cdot \boldsymbol{U}, \tag{3b}$$

where $\boldsymbol{\Sigma}$ is the stress amplitude tensor; $\boldsymbol{U}$ is the displacement amplitude vector; $\boldsymbol{S}$ is a third-order tensor; the effective mass density tensor $\boldsymbol{\rho}_{\mathit{eff}}$ is an explicit function of angular frequency $\omega$. These equations are form-invariant. In addition, all parameters inside are local if the microstructure of the material is sufficiently small compared with the interested wave length (Milton et al., 2006). Norris and Shuvalov (2011) further pointed out that the elasticity tensor $\boldsymbol{C}$ is generally non-symmetric when other gauges are adopted in the transformation.

However, according to *the principle of general invariance*, all laws of physics must be invariant under general coordinate transformations (Ohanian and Ruffini, 2013). I.e., form-invariance is an intrinsic property of correct wave equations, which should be expressed in tensorial forms. The requirement of *the principle of material objectivity* (Truesdell and Noll, 2004) in continuum mechanics could be regarded as a special case of *the principle of general invariance*, when the transformation is limited to an orthogonal mapping. According to these principles, Xiang (2014) pointed out that the proof of the form-invariance of wave equations does not need any assumption of the relation between field variables before and after the transformation. As a by-product, one can naturally obtain elastodynamic equations in time domain:

$$\boldsymbol{\sigma} = \boldsymbol{C} : \nabla \boldsymbol{u} + \boldsymbol{S} \cdot \boldsymbol{u}, \tag{4a}$$

$$\nabla \cdot \boldsymbol{\sigma} + \boldsymbol{f}^a = \boldsymbol{S}^{\mathrm{T}} : \nabla \boldsymbol{u} + \boldsymbol{K} \cdot \boldsymbol{u} + \boldsymbol{\rho} \cdot \ddot{\boldsymbol{u}}, \qquad (4b)$$

where $\boldsymbol{\rho}$ is the mass density tensor; and $\boldsymbol{f}^a$ is a body force vector. The above equations are very similar to the Willis equations, if $\boldsymbol{C}$ is symmetric and $\boldsymbol{K} = \boldsymbol{0}$. In a special case when taking the deformation gradient as the gauge between the displacements before and after the transformation, $\boldsymbol{C}$ is symmetric and the frequency version of Eq. (4) is exactly the same as Eq. (3) if ignoring the body force $\boldsymbol{f}^a$.

Based on the dimensional analysis of Eq. (4a), $\boldsymbol{S}$ can be intuitively regarded as the gradient of the pre-stress (Xiang, 2014). Since $\boldsymbol{K}$ has some relations with $\boldsymbol{S}$, it should have some relations with the gradient of the pre-stress. In addition, according to *the second law of thermodynamics*, an adiabatically separated system will become homogeneous eventually. Therefore, inhomogeneous materials must be the results of certain external forces, which may introduce the inside pre-stress. In this sense, the pre-stress is the concomitant of inhomogeneity. For example, all materials are strictly inhomogeneous due to the existence of interfaces or surfaces. Crystals near the interface or the surface are different from the crystals in other parts and consequently one can find interface or surface stresses (Müller and Saúl, 2004). And it is natural to read some reports about elastic cloaking by employing the pre-stress (Parnell, 2012; Norris and Parnell, 2012; Colquitt et al., 2014).

This paper gives a rigorous proof that the presence of inhomogeneous pre-stresses can result in the Willis equations, in which $\boldsymbol{S}$ is the gradient of the pre-stress and $\boldsymbol{K}$ is related with the effective body force due to the gradient of the pre-stress. With these clear physical meanings, one can find many important applications of the Willis equations.

## 2. The constitutive equation

As shown in Fig. 1, an initial configuration $B^0$ contains inhomogeneous pre-stresses $\sigma^0$, which can be induced by arbitrary processes, such as the elastic deformation, the plastic deformation, the phase transition, etc. A material point in $B^0$ is labeled by its position vector $x$ in the global Cartesian coordinate system with the basis of $e_1$, $e_2$ and $e_3$. Under a small perturbation, $B^0$ moves to an adjacent configuration $B^1$. Consequently, that material point moves to a new spatial position $x'$ in $B^1$ with the displacement $u = x' - x$.

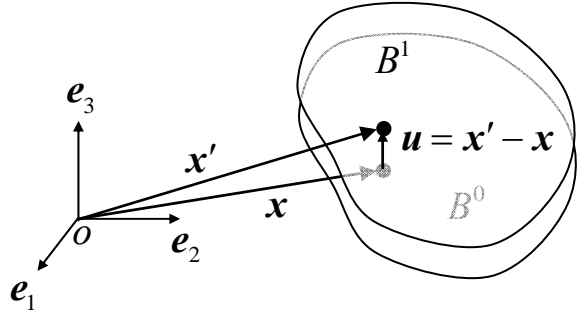

**Fig. 1.** The incremental model.

The constitutive equation describes the relation between the incremental Cauchy stress $\sigma$ at the material point $x$ in $B^0$ and its deformation $u$. To this end, one can firstly consider the Cauchy stress of this material point at the spatial position $x'$ in $B^1$, which is denoted as $\sigma^1(x')$ and can be expanded around the spatial position $x$ as:

$$\sigma^1(x') \approx \sigma^1(x) + \left[\sigma^1(x)\nabla\right] \cdot u, \tag{5}$$

where

$$\sigma^1(x) = \sigma'(x) + \bar{\sigma}(x). \tag{6}$$

$\sigma'(x)$ in Eq. (6) is the Cauchy stress at the spatial position $x$ due to the pure rigid translation $\bar{u}$ during the perturbation from $B^0$ to $B^1$. In this circumstances, the material point at the spatial position $x - \bar{u}$ in $B^0$ moves to the spatial position $x$ in $B^1$. Therefore,

$\sigma'(x)$ in $B^1$ equals to the Cauchy stress at the spatial position $x - \bar{u}$ in $B^0$:

$$\sigma'(x) = \sigma^0(x - \bar{u}) \approx \sigma^0(x) - \left[\sigma^0(x)\nabla\right] \cdot \bar{u}. \tag{7}$$

It is clear that $\sigma'(x) = \sigma^0(x)$ when the rigid translation is constrained ($\bar{u} = 0$) or the pre-stress is homogeneous ($\sigma^0(x)\nabla = 0$).

$\bar{\sigma}(x)$ in Eq. (6) denotes the incremental Cauchy stress at the spatial position $x$ due to the stretch and the rigid rotation during the perturbation from $B^0$ to $B^1$. It is well-known that

$$\nabla u = \varepsilon + \Omega, \tag{8}$$

where the strain $\varepsilon = \frac{1}{2}(\nabla u + u\nabla)$ is a symmetric tensor that represents the stretch; and $\Omega = \frac{1}{2}(\nabla u - u\nabla)$ is an antisymmetric tensor that represents the rigid rotation. If the elasticity tensor $C$, which represents the effective tangent stiffness, is symmetric, the corresponding incremental Cauchy stress is

$$\bar{\sigma}(x) = C : \nabla u = C : \varepsilon. \tag{9}$$

Substituting Eqs. (7) and (9) into Eq. (6), yields:

$$\sigma^1(x) \approx \sigma^0(x) + C : \nabla u - \left[\sigma^0(x)\nabla\right] \cdot \bar{u}. \tag{10}$$

Thus, one can further obtain the low order linear approximation:

$$\left[\sigma^1(x)\nabla\right] \cdot u \approx \left[\sigma^0(x)\nabla\right] \cdot u. \tag{11}$$

Substituting Eqs. (10) and (11) into Eq. (5), obtains the incremental Cauchy stress at that material point as:

$$\sigma(x) = \sigma^1(x') - \sigma^0(x) \approx C : \nabla u + \left[\sigma^0(x)\nabla\right] \cdot (u - \bar{u}). \tag{12}$$

Compared with Eq. (4a), it is clear that $S = \sigma^0\nabla$, the gradient of the pre-stress. In addition, neither the rigid translation nor the rigid rotation can introduce the incremental Cauchy stress according to Eq. (12).

## 3. The equation of motion

### 3.1. The general formulation

The equation of motion at the spatial position $x'$ in $B^1$ is:

$$\nabla' \cdot \sigma^1(x') + f^1 = \rho \cdot \ddot{u}, \quad (13)$$

where $f^1$ is the body force in $B^1$; and the effective mass density $\rho$ is in tensorial form. The heterogeneous property of mass density exists in heterogeneous material at a certain wave frequency. This was demonstrated rigorously by Willis (1985) and can be understood by using the model of heterogeneous microstructures at local resonance (Milton and Willis, 2007).

In index notation, $\nabla' \cdot \sigma^1(x')$ can be written as

$$\begin{aligned}
\frac{\partial \sigma_{ij}^1}{\partial x_i'} &= \frac{\partial x_k}{\partial x_i'} \frac{\partial \sigma_{ij}^1}{\partial x_k} \\
&= \frac{\partial}{\partial x_i'}(x_k' - u_k) \frac{\partial}{\partial x_k}(\sigma_{ij}^0 + \sigma_{ij}) \\
&= (\delta_{ki} - u_{k,i}) \frac{\partial}{\partial x_k}(\sigma_{ij}^0 + \sigma_{ij}) \\
&= \frac{\partial \sigma_{ij}}{\partial x_i} + \frac{\partial \sigma_{ij}^0}{\partial x_i} - u_{k,i} \frac{\partial \sigma_{ij}^0}{\partial x_k} - u_{k,i} \frac{\partial \sigma_{ij}}{\partial x_k}
\end{aligned} \quad (14)$$

Since $u_{k,i} \dfrac{\partial \sigma_{ij}}{\partial x_k}$ is a high order small term, it can be neglected to obtain a linear expression:

$$\nabla' \cdot \sigma^1 \approx \nabla \cdot \sigma + \nabla \cdot \sigma^0 - (\sigma^0 \nabla)^T : \nabla u. \quad (15)$$

The equilibrium equation at the spatial position $x$ in $B^0$ is:

$$\nabla \cdot \sigma^0(x) + f^0 = 0, \quad (16)$$

where $f^0$ is the body force in $B^0$.

Denoting the incremental body force as $f = f^1 - f^0$, and substituting Eqs. (15) and (16) into Eq. (13), obtains:

$$\nabla \cdot \sigma + f \approx (\sigma^0 \nabla)^T : \nabla u + \rho \cdot \ddot{u}. \quad (17)$$

This is the equation of motion in the Willis form.

### 3.2. The special formulation corresponding to the transformation elastodynamics

Eq. (17) does not contain the term $K$ in Eq. (4b), which is also implicitly incorporated into the $\rho_{eff}$ in Eq. (3b) (Milton et al., 2006). Recalling that $K$ appears only when the transformation elastodynamics is applied (Milton et al., 2006; Xiang, 2014), it can be regarded as the result of this special process.

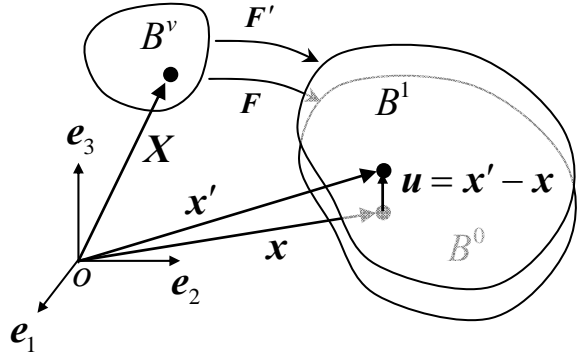

**Fig. 2.** The virtual space $B^v$ is transformed into $B^0$ and $B^1$ under the deformation gradient $F$ and $F'$, respectively.

The virtual space in the transformation elastodynamics is usually an isotropic and homogeneous configuration, which is denoted as $B^v$ in Fig. 2. A material point at spatial position $X$ in $B^v$ is transformed into $x$ in $B^0$ and $x'$ in $B^1$ under the deformation gradient $F$ (with components $F_{i\alpha} = \partial x_i / \partial X_\alpha$) and $F'$ (with components $F'_{i\alpha} = \partial x'_i / \partial X_\alpha$), respectively. In these cases, one can have the relation:

$$\boldsymbol{f}^1 = \frac{J}{J'} \boldsymbol{f}^0, \tag{18}$$

where $J = \det \boldsymbol{F}$ and $J' = \det \boldsymbol{F'}$, which measure the change in material volumes. Under the small deformation, $J' \approx J + (J\nabla) \cdot \boldsymbol{u}$. Therefore, $\boldsymbol{f}^1 \approx \left[1 - J^{-1}(J\nabla) \cdot \boldsymbol{u}\right] \boldsymbol{f}^0$ and

$$\boldsymbol{f} \approx -\left(J^{-1}\boldsymbol{f}^0 \otimes J\nabla\right) \cdot \boldsymbol{u}. \tag{19}$$

Substituting Eq. (19) into Eq. (17), obtains:

$$\nabla \cdot \boldsymbol{\sigma} \approx \left(\boldsymbol{\sigma}^0 \nabla\right)^{\mathrm{T}} : \nabla \boldsymbol{u} + \left(J^{-1}\boldsymbol{f}^0 \otimes J\nabla\right) \cdot \boldsymbol{u} + \boldsymbol{\rho} \cdot \ddot{\boldsymbol{u}}. \tag{20}$$

Compared Eq. (20) with Eq. (4b), it is clear that $\boldsymbol{K} = J^{-1}\boldsymbol{f}^0 \otimes J\nabla$ or $\boldsymbol{K} = -J^{-1}\boldsymbol{\sigma}^0\nabla \otimes J\nabla$ referring to Eq. (16). In this circumstances, $\boldsymbol{f}^0$ can be regarded as the driving body force that transforms $B^v$ to $B^0$. And the $\boldsymbol{f}^a$ in Eq. (4b) could be regarded as an additional body force during the perturbation from $B^0$ to $B^1$.

## 4. Discussions

*4.1. The influence of the gradient of the pre-stress*

The proofs in Section 2 and Section 3 support the intuitive understanding of Xiang (2014) that $S$ and $K$ in Eq. (4) are related with the gradient of the pre-stress.

Substituting Eq. (1a) into Eq. (1b), one can obtain the Lamé-Navier equation

$$\nabla \cdot (C : \nabla u) + f = \rho \ddot{u}. \tag{21}$$

While substituting Eq. (12) into Eq. (17), yields

$$\begin{aligned}
\nabla \cdot (C : \nabla u) + f &\approx \rho \cdot \ddot{u} + (\sigma^0 \nabla)^{\mathrm{T}} : \nabla u - \nabla \cdot \left[ (\sigma^0 \nabla) \cdot (u - \bar{u}) \right] \\
&= \rho \cdot \ddot{u} - \left[ (\nabla \cdot \sigma^0) \nabla \right] \cdot (u - \bar{u}) \\
&= \rho \cdot \ddot{u} + (f^0 \nabla) \cdot (u - \bar{u})
\end{aligned} \tag{22}$$

Eq. (22) has two differences from the Lamé-Navier equation. One is the tensorial mass density $\rho$, which is natural for heterogeneous media (Willis, 1985; Milton and Willis, 2007). Another is the additional term $\left[ (\nabla \cdot \sigma^0) \nabla \right] \cdot (u - \bar{u}) = -(f^0 \nabla) \cdot (u - \bar{u})$, which can be regarded as the effective body force due to the gradient of the pre-stress. If $f^0 \nabla = 0$, Eq. (22) is almost the same as the Lamé-Navier equation, except for the tensorial mass density $\rho$.

It has been reported that the difference between the Willis equations and the classical elastodynamic equations is smaller when the wave frequency is higher (Hu et al., 2011; Xiang, 2014). The reason is probably due to the decoupling of the transverse wave and the longitudinal wave at high frequencies (Cerveny, 2001). In that case, one can use eikonal equation to describe the ray behavior of the elastic wave. So what will happen when the wave frequency is lower and lower and the wave finally approaches to the quasi-static response? To answer this question, one can go over Section 2 and Section 3 of this paper and find out that the deduction of the constitutive equation has nothing to do with the inertia, and the equation of motion (17) is also valid for quasi-static deformation if neglecting the inertia. This means the Willis equations are also valid for quasi-static problems if the inertia can be neglected. I.e., the Willis equations could be more generally valid and a crucial part is the influence of the gradient of the pre-stress, which is the concomitant of inhomogeneity.

*4.2. Potential applications of the Willis equations*

With the presence of pre-stresses one could find many applications of the Willis equations, since their physical meanings are clear.

The straight forward application is in seismology, where the earth crust is inhomogeneous and the pre-stress inside must be balanced with the gravity force. In this case, neglecting the effective body force $\left[\left(\nabla\cdot\boldsymbol{\sigma}^0\right)\nabla\right]\cdot\left(\boldsymbol{u}-\bar{\boldsymbol{u}}\right) = -\left(\boldsymbol{f}^0\nabla\right)\cdot\left(\boldsymbol{u}-\bar{\boldsymbol{u}}\right)$ could introduce some biases in predictions.

Another impact of the Willis equations could be on the physics of interfaces, e.g., the interface between the fiber (or the grain) and the matrix in composites, the crack tip and the surface (Müller and Saúl, 2004), etc., where the gradient of the pre-stress cannot be neglected.

Since the gradient of the pre-stress plays an important role in the Willis equations, the material length scale is thus naturally considered. As mentioned above, the pre-stress near the surface has non-neglectable gradient. Therefore, the material properties near the surface are naturally various in depths according to Eq. (12). Consequently, the effective material properties of small components could be very size sensitive.

And the Willis equations also shed a light on the design of elastic metamaterials. I.e., the trajectory of elastic wave could be controlled by properly distributed pre-stresses (Xiang, 2014).

## 5. Conclusions

With the presence of pre-stresses, this paper demonstrates rigorously that the differences between the Willis equations and the classical elastodynamic equations are mainly due to the gradient of the pre-stress. This is a reason why the Willis equations can give more accurate prediction for inhomogeneous media, if noticing that the pre-stress is the concomitant of inhomogeneity. More confidence on the validity of the Willis equations comes from their compliance with the principle of general invariance. And many exciting applications based on the Willis equations could be expected.

**Acknowledgment**

The authors are grateful for the support from the National Science Foundation of China through grant 11272168.

**References**

Cerveny V. Seismic Ray Theory. Cambridge: Cambridge University Press, 2001: 2.

Chen H, Chan CT. Acoustic cloaking and transformation acoustics. J. Phys. D: Appl. Phys., 2010, 43: 113001.

Colquitt DJ, Brun M, Gei M, Movchan AB, Movchan NV, Jones IS. Transformation elastodynamics and cloaking for flexural waves. Journal of the Mechanics and Physics of Solids, 2014, 72: 131-143.

Hu J, Chang Z, Hu GK. Approximate method for controlling solid elastic waves by transformation media. Phys. Rev. B., 2011, 84: 201101(R).

Leonhardt U. Optical conformal mapping. Science, 2006, 312: 1777-1780.

Milton GW, Briane M, Willis JR. On cloaking for elasticity and physical equations with a transformation invariant form. New J. Phys., 2006, 8: 248.

Milton GW, Willis JR. On modifications of Newton's second law and linear continuum elastodynamics. Proc. R. Soc. A, 2007, 463: 855-880.

Müller P, Saúl A. Elastic effects on surface physics. Surface Science Reports, 2004, 54: 157-258.

Nassar H, He QC, Auffray N. Willis elastodynamic homogenization theory revisited for periodic media. Journal of the Mechanics and Physics of Solids, 2015, 77: 158-178.

Norris AN, Shuvalov AL. Elastic cloaking theory. Wave Motion, 2011, 48: 525-538.

Norris AN, Shuvalov AL, Kutsenko AA. Analytical formulation of three-dimensional dynamic homogenization for periodic elastic systems. Proc. R. Soc. A., 2012, 468: 1629-1651

Norris AN, Parnell WJ. Hyperelastic cloaking theory: transformation elasticity with pre-stressed solids. Proc. R. Soc. A, 2012, 468: 2881-2903.

Ohanian HC, Ruffini R. Gravitation and space time, 3rd ed. Cambridge: Cambridge University Press, 2013: 279.

Parnell WJ. Nonlinear pre-stress for cloaking from antiplane elastic waves. Proc. R. Soc. A, 2012, 468: 563-580.

Pendry JB, Schurig D, Smith DR. Controlling electromagnetic fields. Science, 2006, 312: 1780-1782.

Shamonina E, Solymar L. Metamaterials: How the subject started. Metamaterials, 2007, 1: 12-18.

Srivastava A, Nemat-Nasser S. Overall dynamic properties of three-dimensional periodic elastic composites. Proc. R. Soc. A., 2012, 468: 269-287.

Truesdell C, Noll W. The non-linear field theories of mechanics 3rd Ed. Heideberg: Springer, 2004: 44-47.

Willis JR. Variational principles for dynamics problems in inhomogeneous elastic media. Wave Motion, 1981, 3: 1-11.

Willis JR. The nonlocal influence of density variations in a composite. Int. J. Solids Structures, 1985, 7: 805-817.

Willis JR. Dynamics of composites continuum micromechanics. In Suquet P, eds. Continuum micromechanics: CISM Courses and Lectures No. 377. Berlin: Springer-Verlag, 1997: 265-290.

Willis JR. Effective constitutive relations for waves in composites and metamaterials. Proc. R. Soc. A., 2011, 467: 1865-1879.

Xiang ZH. The form-invariance of wave equations without requiring a priori relations between field variables. SCIENCE CHINA Physics, Mechanics & Astronomy, 2014, 57: 2285-2296.